\documentstyle[aps,prl,twocolumn,epsf]{revtex}
\begin{document}
\title{A new mechanism of neutron star radiation}
\author{Anatoly A.~Svidzinsky}
\address{Bartol Research Institute, University of Delaware, Newark, DE 19716,
USA}
\date{\today }
\maketitle

\begin{abstract}
We find a new mechanism of neutron star radiation wherein radiation is
produced by the stellar interior. The source of radiation is oscillating
neutron vortices in the superfluid core of a rotating neutron star.
Thermally excited helical waves of vortices generate fast magnetosonic waves
at the stellar crust. Near the crust bottom such waves reduce to a
collisionless zero sound in an electron liquid, while near the stellar
surface they behave as electromagnetic waves in a medium. The magnetosonic
waves propagate across the crust and transform into electromagnetic
radiation at the surface. The vortex contribution has the spectral index $%
\alpha \approx -0.45$ and can explain nonthermal radiation of middle-aged
pulsars observed in infrared, optical and hard X-ray bands. Detection of
vortex radiation allows direct determination of the core temperature.
Comparing the theory with available spectra observations we find that the
core temperature of the Vela pulsar is $T\approx 8\times 10^8$K, while the
core temperature of PSR B0656+14 exceeds $2\times 10^8$K. This is the first
measurement of the temperature of a neutron star core. The temperature
estimate rules out equation of states incorporating Bose condensations of
pions or kaons and quark matter in these objects. In principle, zero sound
can also be emitted by other mechanisms, rather than vortices, which opens a
perspective of direct spectroscopic study of superdense matter in the
neutron star interiors.
\end{abstract}


Properties of matter at densities much larger than nuclear are poorly known
and constitute a challenging problem of modern science. They have broad
implications of great importance for cosmology, the early universe, its
evolution, for compact stars and for laboratory physics of high-energy
nuclear collisions. Present searches of matter properties at high density
take place in several arenas. One of them is investigation of neutron stars.

Neutron stars (NSs) are compact objects of radius about $10$km and the mass
of the order of solar mass. Cores of NSs consist of superfluid neutrons with
some (several per cent by particle number) admixture of protons and
electrons. More dense NSs also possess an inner core. Its radius may reach
several kilometers and central density can be much larger then the nuclei
density. Several hypotheses about composition and equation of state of the
inner core are discussed in the literature. One of them is appearance of $%
\Sigma $- and $\Lambda $-hyperons. The second hypothesis assumes the
appearance of pion or kaon condensation. The third hypothesis predicts a
phase transition to strange quark matter composed of almost free $u$, $d$
and $s$ quarks with a small admixture of electrons.

The cooling rate and temperature of the central part of a NS substantially
depends on the properties of dense matter. In this paper we find that NS
radiation can be produced in its interior, rather then at the surface.
Detection of such radiation in the $X-$ray band allows direct determination
of the core temperature. We find that the core temperature of the Vela
pulsar is $T\approx 8\times 10^8$K, while the core temperature of PSR
B0656+14 exceeds $2\times 10^8$K. This estimate excludes exotic equation of
states incorporating Bose condensations of pions or kaons and quark matter
in these objects. A systematic study of pulsar radiation in the $X-$ray band
can locate objects with fast cooling core which could be candidates for NSs
with exotic states of matter.

{\it Zero sound in electron liquid} --- For middle aged NSs the effective
frequencies of electron-electron and electron-phonon collisions in the
stellar crust are less than $10^{12}$Hz. In the core the collisions of
electrons with baryons are suppressed by neutron superfluidity and proton
superconductivity. As a result, at large frequencies the electrons in the
stellar interior can be treated as an independent system of Fermi particles
which move in collisionless regime. In such regime a zero sound is a
possible collective excitation of Fermi liquid \cite{Land57}. Speed of zero
sound $u_s$ is larger than the electron Fermi velocity $v_F$. In the
presence of magnetic field $H$ equations of electron motion reduce to
equations of magnetic hydrodynamics in which $u_s$ enters the equations
instead of speed of usual sound and the Alfven velocity $u_A$ is estimated
in terms of the electron density $\rho _e$: $u_A=H/\sqrt{4\pi \rho _e}$.
Magnetic field at the stellar crust couples electron sound and
electromagnetic wave into one excitation known as a fast magnetosonic wave
(FMW). Near the crust bottom $u_A\ll u_s\approx c$, where $c$ is the speed
of light, and the FMWs reduce to zero sound. However, near the stellar
surface (at small density) $u_A>u_s$ and the FMWs behave as electromagnetic
waves in a medium with dielectric constant $\varepsilon =1+c^2/u_A^2$. As a
result, at the stellar surface the FMWs transform into electromagnetic
radiation in vacuum the same way as refraction of usual electromagnetic
waves at the dielectric-vacuum interface. Typically near the stellar surface
$u_A\gg c$ and, hence, the transmission coefficient is close to $1$.

The key idea of the paper is that NS interior is transparent for zero sound
(FMWs). We assume that the crust matter is in a crystal phase: $%
T<T_m=2.5\times 10^7\rho _6^{1/3}$K, where $T_m$ is the melting temperature,
$\rho _6$ is the matter density in units $10^6{\rm g/cm}^3$ \cite{Hern84}.
Near the stellar surface there is a phase transition between a gaseous
atmosphere and a condensed metallic phase and the atmosphere has negligible
optical depth \cite{Lai01}. The density of the metallic phase at the surface
depends on magnetic field. E.g., for iron crust and $H=10^{12}{\rm Gs}$ the
surface density is $\rho \approx 10^3{\rm g/cm}^3$ \cite{Thor98}.

For a weakly interacting Fermi gas, such as a gas of electrons in the NS
interior, the attenuation of zero sound is exponentially small \cite{Gott60}%
. However, in the presence of magnetic field the FMWs are accompanied by
oscillation of electric field which results in partial damping of FMWs
near the stellar surface. The attenuation increases with decreasing the
matter density. However, at $\rho _6\lesssim 0.6H_{12}^{3/2}$ the system as
a whole occupies only a small number of Landau levels \cite{Hern84}. At such
densities the electron motion becomes effectively one dimensional and the
electron scattering, as well as magnetosonic attenuation, is suppressed by
magnetic field. The maximum damping occurs in a boundary region where
magnetic field still does not suppress the electron scattering. FMWs pass
the critical region and reach the surface without attenuation if in the
region
\begin{equation}
\label{d242}\left| \sin \theta \right| <0.004H_{12}^2g_{15}^{1/2}/T_6^{5/2},
\end{equation}
here $\theta $ is the angle between the wavevector and the magnetic field, $%
g_{15}$ is the acceleration of gravity in units $10^{15}{\rm cm/s}^2$ \cite
{Svid02}. For $T=10^6$K, $H=4\times 10^{12}{\rm Gs}$ and $g=10^{15}{\rm cm/s}%
^2$ only waves with $\left| \theta \right| <3.7^{\circ }$ reach the stellar
surface. Below we show that zero sound in NS interior is generated at the
core-crust interface by superfluid neutron vortices which undergo thermal
helical oscillations. Vortices themselves are excited by scattering of
single-particle excitations on the vortex cores. Sound waves propagate
across the stellar crust towards the surface. Due to anisotropic attenuation
in the surface region and star rotation the observed vortex radiation is
pulsed. According to Eq. (\ref{d242}), for colder NSs the shape of vortex
pulses becomes broader.

{\it Vortices in neutron stars} --- If a NS rotates with an angular velocity
${\bf \Omega =}\Omega \hat z$ a periodic vortex lattice forms in a
superfluid neutron phase of the stellar core. The characteristic size of the
vortex is $\xi =10^{-12}{\rm cm}$ and the total number of vortices for $%
\Omega /2\pi =10{\rm s}^{-1}$ is $N=2m_nR^2\Omega /\hbar \approx 10^{17}$,
where $m_n$ is the neutron mass and $R\approx 10{\rm km}$ is the radius of
the superfluid. Let us consider a single straight neutron vortex line with
length $L$. Normal modes of the vortex are helical waves \cite{Lifs80}
\begin{equation}
\label{q1}r=A\exp (i\omega t+ik_{\parallel }z),\quad \omega =\hbar
\,k_{\parallel }^2\ln (1/k_{\parallel }\xi )/4m_n,
\end{equation}
where $A$ is the amplitude and $k_{\parallel }$ is the wave number. Eqs. (%
\ref{q1}) are valid for $k_{\parallel }\xi \ll 1$, or $\omega \ll \omega
_c=\hbar /4m_n\xi ^2\approx 10^{20}$Hz and describe rotation of a helix with
the angular frequency $\omega $. At distances $r\lesssim \xi $ (we use
cylindrical coordinates $r,\phi ,z$) the helical vortex motion can be
described in terms of the perturbation in the velocity potential $\Phi
^{\prime }$
\begin{equation}
\label{q2}\Phi ^{\prime }=iA\hbar \exp \left[ i(\omega t+k_{\parallel
}z-\phi )\right] /2m_nr.
\end{equation}
In the bulk the function $\Phi ^{\prime }$ should satisfy a linearized
equation of superfluid hydrodynamics, which is, in fact, the wave equation
of sound propagation
\begin{equation}
\label{q3}\partial ^2\Phi ^{\prime }/\partial t^2-c_s^2\nabla ^2\Phi
^{\prime }=0,
\end{equation}
where $c_s$ is the speed of sound in superfluid. In the limit $\omega \ll
\omega _c$ one can omit the first term in Eq. (\ref{q3}). Then solution of
this equation which satisfies the boundary condition (\ref{q2}) has the form
\begin{equation}
\label{q4}\Phi ^{\prime }=iA\hbar k_{\parallel }K_1(k_{\parallel }r)\exp
\left[ i(\omega t+k_{\parallel }z-\phi )\right] /2m_n,
\end{equation}
where $K_1$ is the modified Bessel function. The solution exponentially
decreases at $r\gg 1/k_{\parallel }$, that is perturbation in the superfluid
velocity ${\bf V}=\nabla \Phi $ is localized near the vortex core and no
energy is radiated.

{\it Generation of zero sound at the core boundary} --- The bulk of
superconducting protons in NS core do not rotate by forming vortices \cite
{Alpa84}. However, due to interactions between protons and neutrons, neutron
superfluid velocity generates a superfluid current of protons in the
vicinity of neutron vortices (drag effect). At the interface between the
stellar crust and the outer core the matter undergoes a first order
transition from an $Aen$ phase to a uniform liquid of neutrons, protons and
electrons ($npe$ phase). Nuclei in the $Aen$ phase form a Coulomb crystal.
In the $npe$ phase the proton liquid is superconducting, while in the $Aen$
phase protons constitute a part of nuclei. According to Eq. (\ref{q4}),
helical vortex motion does not generate sound waves inside the stellar core.
However, such waves are generated at the interface between the
superconducting and $Aen$ phases.

Due to the drag effect helical waves of a neutron vortex produce density
oscillation of protons in the superconducting phase. Free electrons screen
the electric field and oscillate together with protons. In $Aen$ phase there
is no drag effect and electron motion is not coupled with neutron vortices.
To estimate the power of radiated zero sound we note that kinetic energy of
ultrarelativistic electrons at the crust bottom is much larger than their
interaction energy. Hence, the radiated power is approximately the same as
in the limiting case of non interacting electrons. In this limit, the
electrons move in a ballistic regime and the energy flux across the
crust-core interface is given by the surface integral $Q=\int E_ev_FdS,$
where $v_F\approx c$ and $E_e$ is the contribution to the electron energy
density due to the helical vortex motion. Density oscillation of electrons
(which is accompanied by large change in their Fermi energy) determine the
helix energy in the stellar core which is equal to $\int E_eLdS$. As a
result, the average energy flux of sound waves produced by a helical wave of
a single vortex is
\begin{equation}
\label{v5}Q=\frac cL\int E_eLdS=\frac cL\frac{\hbar \omega }{[\exp (\hbar
\omega /k_BT)-1]},
\end{equation}
where $T$ is the core temperature. One should note that the wave length of
sound is much larger than the size of the area from which it is generated
and, hence, the sound waves have approximately spherical front. Taking into
account that the number of vortex modes within the interval $dk_{\parallel }$
is $Ldk_{\parallel }/2\pi $ and the number of vortices $N=2m_nR^2\Omega
/\hbar $, we obtain the spectral density of sound waves power radiated from
the oscillating vortex lattice in a half space
\begin{equation}
\label{mss}P_{{\rm v}}(\omega )=\frac{\sqrt{2}cm_n^{3/2}R^2\Omega \sqrt{%
\omega }}{\pi \hbar ^{3/2}\sqrt{\hbar \ln (\omega _c/\omega )}\left[ \exp
(\hbar \omega /k_BT)-1\right] }.
\end{equation}
Fig. 1 shows vortex mechanism of NS radiation.

\begin{figure}
\bigskip
\centerline{\epsfxsize=0.36\textwidth\epsfysize=0.50\textwidth
\epsfbox{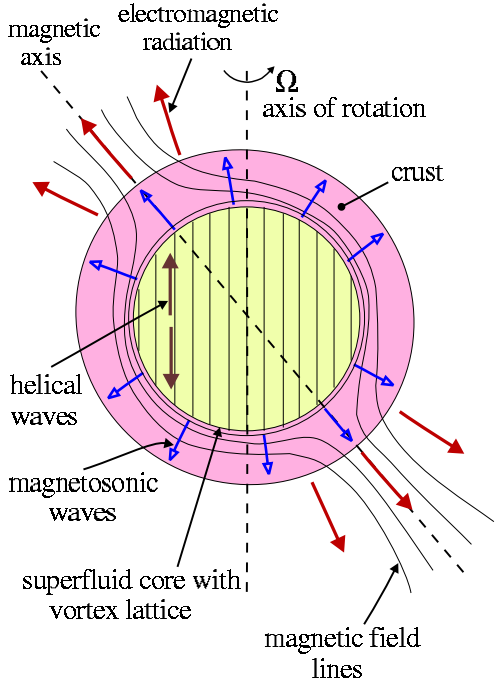}}

\vspace{0.6cm}

\caption{Mechanism of a NS radiation. Thermaly excited helical waves
of neutron vortices in the superfluid core produce FMWs in the
stellar crust. FMWs propagate across the crust and transform into
electromagnetic radiation at the star surface. Mainly the radiation comes out
from regions with strong magnetic field (near magnetic poles).}
\label{fig1}
\end{figure}

One should note that electromagnetic radiation produced by vortices can not
exceed black body radiation produced by the surface $4\pi R^2$ with the same
temperature $T$ (Kirchhoff's law). We assume that vortices are excited
thermally and if at some frequency the vortex radiation becomes comparable
with radiation of a black body this means that at such frequency the rate of
thermal excitation imposes the main restriction on the radiated power:
vortices radiate maximum power which can be pumped from the thermal
reservoir. Kirchhoff's law results in the following limitation at which Eq. (%
\ref{mss}) describes vortex radiation
\begin{equation}
\label{m102}\omega /2\pi >0.24\hbar ^{-3/5}c^{6/5}m_n^{3/5}\Omega ^{2/5}\ln
^{-1/5}(\omega _c/\omega ).
\end{equation}
For $\Omega /2\pi =4{\rm s}^{-1}$ we obtain $\omega /2\pi >1.5\times 10^{14}$%
Hz. At lower frequencies the spectrum of vortex radiation follows Planck's
formula.

To compare our theory with observations it is convenient to represent the
spectral density of vortex radiation $P_{{\rm v}}(\omega )$ as a power law $%
P_{{\rm v}}(\omega )\propto \omega ^\alpha $, where $\alpha $ is the
spectral index. The $\sqrt{\ln (\omega _c/\omega )}$ function in Eq. (\ref
{mss}) shifts the spectral index by a small value $1/2\ln (\omega _c/\omega
) $. The shift depends weakly on $\omega $ and changes the spectral index
from $\alpha \approx -0.46$ in the optical band to $\alpha \approx -0.43$ in
the $X-$ray band. Apart from vortex contribution there is thermal radiation
from the NS surface. We estimate thermal radiation assuming helium
atmosphere and use numerical results obtained by Romani \cite{Roma87}. The
model fits well the thermal spectrum with only one parameter, the effective
surface temperature $T_{{\rm eff}}$. The total radiation from a NS in unit
solid angle is given by the sum of vortex and thermal $P_{{\rm th}}$
components:
\begin{equation}
\label{m103}P(\omega )=aP_{{\rm v}}(\omega )/2\pi +P_{{\rm th}}(\omega ).
\end{equation}
Here we introduced a dimensionless free parameter $a\approx 0.1\div 1$ which
takes into account partial absorption of the vortex radiation near the
stellar surface and geometrical effects related to unknown magnetic field
distribution and position of the line of site.

{\it Discussion} --- Fig. 2 compares the observed radiation spectrum of
middle-aged pulsars PSR B0656+14 ($\Omega /2\pi =2.6{\rm s}^{-1}$) and Vela (%
$\Omega /2\pi =11.2{\rm s}^{-1}$) with the spectrum predicted by our theory.
Thermal radiation of the stellar surface dominates in the ultraviolet and
soft $X-$ray bands. The effective surface temperature is $T_{{\rm eff}%
}=4.9\times 10^5$K for PSR B0656+14 and $T_{{\rm eff}}=7.8\times 10^5$K for
Vela. The NS radius $R_s$ (in ${\rm km}$) is related to the pulsar distance $%
D$ (in ${\rm kpc}$) as $R_s=53\sqrt{1+z}D$ for PSR B0656+14 and $R_s=41\sqrt{%
1+z}D$ for Vela ($z$ is the NS redshift). The vortex contribution (dash
line) dominates in infrared, optical and hard $X-$ray bands. In the far
infrared band the radiation spectrum changes its behavior and follows
Planck's formula. The sum of the vortex and thermal components is displayed
with the solid line. For PSR B0656+14 we take the core temperature $%
T=6.4\times 10^8$K and the geometrical factor $a=0.18$, while for Vela $%
T=8\times 10^8$K and $a=0.33$. The observed broad-band spectrum is
consistent with our model for typical NS parameters, which suggests that the
vortex mechanism of radiation operates in a broad frequency range from IR to
hard $X-$rays. Also in the optical band the radiation of middle-aged pulsars
was found to be highly pulsed with pulses similar to those in the hard $X-$%
ray band \cite{Shea98,Shea97}. This agrees with the vortex mechanism which
predicts pulsed radiation.

\begin{figure}
\bigskip
\centerline{\epsfxsize=0.33\textwidth\epsfysize=0.37\textwidth
\epsfbox{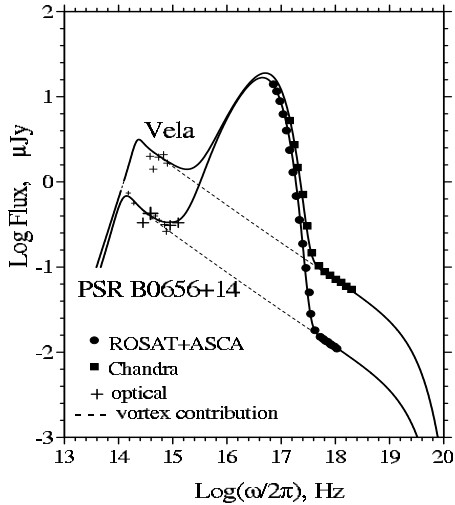}}

\vspace{0.6cm}

\caption{Broadband spectrum of PSR B0656+14 [12] and
the Vela pulsar [13,14]. Solid line
is the fit by the sum of the vortex and thermal components. Thermal radiation of
the stellar surface dominates in the ultraviolet and soft $X-$ray bands, while
the vortex contribution (dash line) prevails in infrared, optical and hard
$X-$ray bands, where its spectrum has a slope $\alpha \approx -0.45$.
In the far infrared band the spectrum changes its
behavior and follows Planck's formula with $P\propto \omega^2$.
}
\label{fig2}
\end{figure}

It is worth to note that the spectral index of vortex radiation is a fixed
parameter in our theory. If a single mechanism of nonthermal radiation
operates in the broad range, then the continuation of the $X-$ray fit to the
optical range determines the spectral index of the nonthermal component with
a big accuracy. Good quantitative agreement of our theory with the observed
spectral index serves as a strong evidence that the vortex mechanism is
responsible for radiation of middle-aged NSs in IR, optical and hard $X-$ray
bands. The vortex contribution exponentially decreases at $\hbar \omega
>k_BT $. Observation of such spectrum behavior allows direct determination
of the core temperature $T$. For PSR B0656+14 and the Vela pulsars the power
law spectrum in the hard $X-$ray band shows no changes up to the highest
frequency $2\times 10^{18}$Hz at which the data are available (see Fig. 2).
This indicates that temperature of the NSs core is larger than $2\times 10^8$%
K. Measurements of spectra of middle-aged pulsars in the $10^{18}-10^{19}$Hz
range are needed to search for possible manifestations of the core
temperature. Recent observation of the Vela pulsar with RXTE has covered the
frequency band $(0.49-7.3)\times 10^{18}$Hz. These data in combination with
OSSE observations, $(1.7-13.8)\times 10^{19}$Hz, allows us to estimate the
temperature of the Vela core. Light curves of the Vela pulsar have several
peaks. We associate the vortex radiation with the second optical peak (see
Fig. 1 in \cite{Hard99} and Fig. 4 in \cite{Stri99}). To estimate the core
temperature one should trace the evolution of the vortex peak in the energy
interval $1$keV$-1$MeV. If we assume that the amplitude of the vortex peak
follows Eq. (\ref{mss}), then we can find total (integrated over frequency)
radiation intensities of vortices in different energy bands. Then we compare
the detected integrated intensities of the vortex peak in the Rossi and OSSE
bands \cite{Stri99} with those predicted by Eq. (\ref{mss}). We found that
if there was no temperature decay of the vortex peak the total vortex
intensity in the OSSE band should be 3 times larger than those actually
observed. This indicates on the decay of the vortex spectrum with the core
temperature $T\approx 8\times 10^8$K.

The estimate of the core temperature allows us to make a conclusion about
the interior constitution of the NSs. A hot NS cools mainly via neutrino
emission from its core. Neutrino emission rates, and hence the core
temperature, depend on the properties of dense matter. If the direct Urca
process for nucleons is allowed such NS cools to $10^8$K in weeks \cite
{Peth92}. However, the characteristic age of PSR B0656+14 is about $10^5$%
yrs, while the Vela pulsar is about $10^4$ years old. So, the NS core cools
down to $10^8$K at least $10^6$ times slower then the rate predicted by the
direct Urca process. This estimate excludes equation of states incorporating
Bose condensations of pions or kaons and quark matter. The point is that the
neutrino emission processes for these states may be regarded as variants of
the direct Urca process for nucleons \cite{Peth92}. As a result, all these
states give rise to neutrino emission, though generally smaller, comparable
to that from the direct Urca process for nucleons which inconsistent with
the discrepancy of the cooling rate in the factor $10^6$.

Detection of vortex radiation opens a possibility to study composition of NS
crust. Since FMWs generated by vortices propagate through the stellar
interior the spectrum of vortex radiation should contain (redshifted)
absorption lines which correspond to low energy excitation of nuclei that
form NS crust. E. g., the $^{57}$Fe nucleus has an excited state with the
energy $14.4$kev$=3.5\times 10^{18}$Hz, which would produce an absorption
line in the hard $X-$ray band if the core temperature is greater than $%
1.7\times 10^8$K. Bottom layers of the crust may contain exotic nuclei with
the mass number up to 600 and the core radiation creates a perspective to
study their properties. Another challenging problem is interaction of zero
sound with exotic states of matter. The point is that generation of zero
sound by vortices is only one of the possible mechanisms. If an exotic state
of matter absorbs zero sound at some frequency, then, according to
Kirchhoff's law, it will radiate zero sound at this frequency. As a result,
spectrum of stellar radiation must contain characteristic emission lines
corresponding to such processes which opens a perspective of direct
spectroscopic study of superdense matter. A detailed version of our theory
is available in \cite{Svid02}.

I am very grateful to A. Fetter, G. Shlyapnikov, M. Binger, S. T. Chui, V.
Ryzhov for valuable discussions and the Aspen Center for Physics where part
of the results has been obtained. This work was supported by NSF, Grant No.
DMR 99-71518 and by NASA, Grant No. NAG8-1427.

\end{document}